\documentclass[twocolumn,aps,showpacs,prb,tightenlines,amsmath,amssymb]{revtex4}
\usepackage{graphicx}    
\usepackage{amssymb}  
\usepackage{dcolumn}
\usepackage{amsmath}     
\usepackage{bm}                                          
\usepackage{colordvi}                   

\begin{document}             
\title{Energy spectra of three electrons in SiGe/Si/SiGe laterally coupled triple quantum dots}
\author{Y. F. Ren}
\affiliation{Hefei National Laboratory for Physical Sciences at Microscale and
  Department of Physics, University of Science and Technology of China, Hefei,
  Anhui, 230026, China} 
\author{L. Wang}
\thanks{wlf@mail.ustc.edu.cn}
 \affiliation{Hefei National Laboratory for Physical Sciences at Microscale and
  Department of Physics, University of Science and Technology of China, Hefei,
  Anhui, 230026, China}
\author{Z. Liu}
 \affiliation{Hefei National Laboratory for Physical Sciences at Microscale and
  Department of Physics, University of Science and Technology of China, Hefei,
  Anhui, 230026, China}
\author{M. W. Wu}     
\thanks{mwwu@ustc.edu.cn}
\affiliation{Hefei National Laboratory for Physical Sciences at Microscale and
  Department of Physics, University of Science and Technology of China, Hefei,
  Anhui, 230026, China} 
\date{\today}

\begin{abstract}
We investigate the energy spectra of three electrons in SiGe/Si/SiGe equilateral
triangular and symmetric linear triple quantum dots in the presence of magnetic
(in either Faraday or Voigt configuration) and electric fields with single
valley approximation by using the real-space configuration interaction method. 
The strong electron-electron Coulomb interaction, which is crucial to the energy
spectra, is explicitly calculated whereas the weak spin-orbit coupling is 
treated perturbatively.
In both equilateral triangular and symmetric linear triple quantum dots, 
we find doublet-quartet transition of ground-state spin configuration by varying 
dot size or interdot distance in the absence of external fields. This 
transition has not been reported in the literature on triple
 quantum dots. In the magnetic-field 
(Faraday configuration) dependence of energy spectra, we find 
anticrossings with large energy 
splittings between the energy levels with the same spin state in the absence of
the spin-orbit coupling. 
This anticrossing behavior originates from the triple quantum dot confinement 
potential. In addition, with the inclusion of the spin-orbit coupling, we find
that all the intersections shown in  the equilateral 
triangular case become anticrossing whereas only part of the intersections in
symmetric linear case show anticrossing behavior in the 
presence of magnetic field in either the Faraday or Voigt configuration. 
All the anticrossing behaviors are
analyzed based on symmetry consideration.
Moreover, we show that the electric field can effectively influence the
energy levels and the charge configurations.

\end{abstract}

\pacs{73.21.La, 73.22.-f, 61.72.uf, 71.70.Ej}
\maketitle

\section{INTRODUCTION}
Spin-based qubits in semiconductor quantum dots (QDs) have been widely 
investigated for promising applications in quantum information processing.
\cite{Loss98,Reimann,Wiel,Hanson,Hawrylak-rev,Marie}
So far, much attention has been paid to spin qubits utilizing single-electron 
Zeeman sublevels and two-electron singlet-triplet states in both single and
double QDs.
\cite{Wiel,Reimann,Hanson,Loss98,Loss99,Levy02,
Hanson03,Taylor05,Taylor05-science,Koppens06,Taylor07,
Foletti,Johnson,Koppens,Borselli,YYW,
lin1,ka1,lin2,culcer10-1,Fabian11,Fabian12,culcer1,Fabian10,
cheng04,jiang10,jiang10-1,Garcia}
Similar to the proposals in single and double QDs, three-electron doublet 
states in linear triple QDs\cite{DiVincenzo00,Taylor10,Taylor13,DiVincenzo12}
(TQDs) or chirality states in triangular TQDs\cite{Hawrylak-b,Gimenez09}
have also been raised to realize spin qubits recently.
Additionally, based on Zeeman sublevels of single electron in each QD 
as a spin qubit, TQDs can be used as a small circuit for constructing 
the QD network.\cite{Acin,Hawrylak-rev,Hawrylak4,Bennett,Roos,Steane} 
This circuit has some intriguing functionalities. Specifically, it 
can be used to realize quantum teleportation without losing information.\cite{Bennett}
Moreover, the entangled Greenberger-Horne-Zeilinger state and Werner 
state in this system can be applied for quantum error
correction.\cite{Roos,Steane} 
All these potential applications may make TQDs attractive
for quantum computation.

Recently, electronic and spin properties of TQDs with few electrons have been
studied both experimentally and theoretically.\cite{Hawrylak-rev, 
Gaudreau06,Gaudreau10,Yamahata,Pierre,Pan,Braakman,
Amaha,
Gaudreau12,Medford13, 
Hawrylak,Hawrylak1,Hawrylak2,Hawrylak1-1,
Scarola,Scarola04,Hawrylak3,Bulka} 
In experiments, the stability diagrams of linear Si TQDs\cite{Pierre,Pan,Yamahata}
and both linear\cite{Gaudreau06,Gaudreau10,Braakman} and triangular\cite{Amaha}
GaAs TQDs have been measured. 
In addition, the coherent manipulation of 
doublet-quartet\cite{Gaudreau12} or doublet-doublet\cite{Medford13} states in
GaAs linear TQDs has been realized very recently. 
Theoretically, both energy spectra and spin configurations of the lowest several
states of few electrons in TQDs have been investigated.\cite{Hawrylak3,Hawrylak,
  Hawrylak1-1,Hawrylak1,Scarola,Scarola04,Hawrylak2,Bulka}
Korkusinsik {\em et al.}\cite{Hawrylak1-1} proposed a set of topological
Hund's rules to understand the spin configurations of the ground states of few
electrons in GaAs TQDs in the absence of external fields.
Delgado {\em et al.}\cite{Hawrylak2} showed the transition of the spin configurations 
of few-electron states in GaAs TQDs by tuning the perpendicular magnetic
field (i.e.,~the Faraday configuration). 
Bu{\l}ka {\em et al.}\cite{Bulka} investigated both the linear and
nonlinear Stark effects induced by the in-plane electric field. 
In addition to these works on triangular TQDs, Hsieh 
{\em et al.}\cite{Hawrylak3} also studied the dependence of the energy spectra on 
detuning energy in linear GaAs TQDs.
It is noted that all the above works are based on the Hubbard model.
Utilizing this model, the electron-electron Coulomb interaction, which is crucial 
to the energy spectra,\cite{lin1,lin2,ka1,Fabian11,Zhe} is not explicitly included, 
but rather given as Hubbard parameters. 
With the electron-electron Coulomb interaction explicitly calculated by
  using the real-space 
configuration interaction method, Hawrylak and Korkusinsik\cite{Hawrylak} investigated the gate-voltage 
dependence of the three-electron energy spectra in triangular GaAs TQDs. However, the effects of the magnetic field and 
spin-orbit coupling (SOC), which have been shown to be very important to the energy spectra, 
were not studied in their work.\cite{lin1,lin2,ka1,Zhe}
It is further noted that most of the theoretical works on electronic and spin
properties of few electrons in TQDs have focused on GaAs until now.
As reported, the spin decoherence, which is essential for applications in quantum 
computation, is limited by the hyperfine interaction\cite{Johnson} and 
spin-orbit coupling (SOC)\cite{Dresselhaus,Rashba1} in GaAs.
In contrast to GaAs, Si has much better spin decoherence properties, which may
make it more attractive.\cite{Taylor05,Vervoort1, Ivchenko,Sarma10,Tahan,Morton} 
However, to the best of our knowledge, theoretical work specific to Si TQDs has
not been reported.

In the present work, we investigate the energy spectra of three electrons in
both equilateral triangular and symmetric linear Si TQDs in the presence of
external magnetic and electric fields.
The strong electron-electron Coulomb interaction is explicitly included by the
real-space configuration interaction method whereas the SOC with much smaller
energy is treated perturbatively. 
We first investigate the case of equilateral triangular TQDs where the
dependences of energy spectra on the external either perpendicular magnetic or
parallel magnetic (i.e., the Voigt configuration) field, dot size, interdot
distance, and electric field are calculated.
We find anticrossings with large energy splittings between the energy levels
with the same spin state in the perpendicular magnetic-field dependence of the
energy spectra in the absence of the SOC. 
These anticrossings, which have not been reported in the literature,
originate from the equilateral triangular TQD confinement potential.
As for the parallel magnetic-field dependence, the energy spectra only vary
linearly due to the negligible orbital effect of the parallel magnetic field. 
In addition, we find doublet-quartet transition of the ground-state spin
configuration by tuning the interdot distance or dot size in the absence of 
external fields and the SOC. We also find that the three-electron energy levels and their 
charge configurations can be strongly affected by the in-plane electric field.
Then we turn to the case of symmetric linear TQDs where the effects of dot size,
interdot distance, magnetic and electric fields on energy spectra are 
discussed. Anticrossings between the energy levels with the same spin
state are observed in the 
perpendicular magnetic-field dependence of the energy spectra, which is similar
to the case of the equilateral triangular TQDs.   
It is noted that with the inclusion of the SOC, all the intersections show anticrossing
behavior in equilateral triangular TQD case whereas only part of them become
anticrossing in symmetric linear TQD case. 

This paper is organized as follows. In Sec.\,II, we introduce the 
model and lay out the formalism. 
Our main results are presented in Sec.\,III where the 
equilateral triangular and symmetric linear TQDs are investigated in Sec.\,IIIA and
Sec.\,IIIB, respectively.
We summarize in Sec.\,IV.

\section{MODEL AND FORMALISM}

\begin{figure}
  \begin{center}
    \includegraphics[width=8cm]{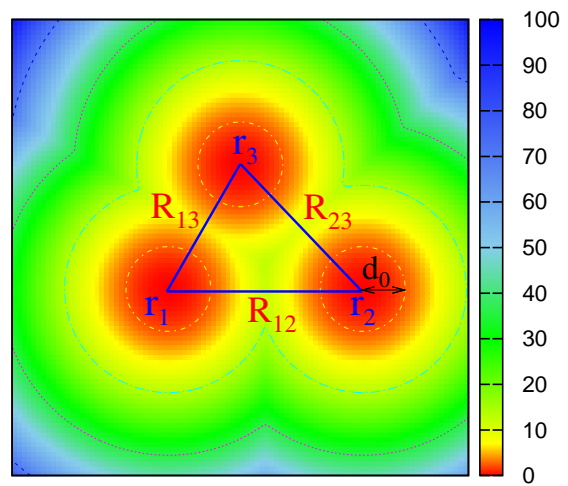}
  \end{center}
  \caption{(Color online) Schematic of the laterally coupled TQD potential. 
    ${\mathbf r}_i$ ($i=1$-$3$) is the location of dot $i$. $d_0$ and $R_{ij}$
    represent the effective dot size and interdot distance between dots $i$ and
    $j$, respectively. It is noted that three dots have same effective
      dot size.}
    \label{fig1}
\end{figure}

We set up our model in a SiGe/Si/SiGe quantum well grown along [001] direction (the $z$-axis).
A strong confinement along this direction splits the six-fold degenerate
conduction band minima or valleys of bulk Si into a four-fold degeneracy and a 
two-fold one with a large energy splitting.\cite{Vrijen,Tahan07} The two-fold
degenerate valleys with lower energy can be further lifted by a valley splitting
in the presence of interface scattering.\cite{Tahan07,Sarma10,Goswami,Ando} 
Here, we focus on a large valley splitting case where only the lowest valley
eigenstate is relevant.\cite{Fabian12,culcer1} Furthermore, we restrict the
system to a two-dimensional one by considering 
that the confinement along the $z$-axis is much stronger than the lateral
ones.\cite{Fabian12,Sarma10}
The lateral confinement potential is chosen as 
$V_{\rm c}({\mathbf r})=\tfrac{1}{2}m_t\omega_0^2{\rm min}\{({\mathbf r}-{\mathbf r}_1)^2,
({\mathbf r}-{\mathbf r}_2)^2,({\mathbf r}-{\mathbf r}_3)^2\}$, where $m_t$ and 
$\omega_0$ represent the in-plane effective mass and confining potential 
frequency, respectively.\cite{Fock1,Darwin1} 
A schematic of TQDs is shown in Fig.~\ref{fig1} where three QDs are located at 
${\mathbf r}_i$ ($i=1$-$3$) with the effective dot size and the
interdot distance between dots $i$ and $j$ being 
$d_0=\sqrt{\hbar\pi/(m_t\omega_0)}$ and $R_{ij}=|{\mathbf r}_i-{\mathbf r}_j|$,
respectively. 
The TQD confinement potential can be tuned by both the dot size and interdot 
distance.
In the present work, we focus on equilateral 
triangular and symmetric linear TQDs with the interdot distances being $R_{12}=R_{23}=R_{13}=x_0$ 
and $R_{13}=R_{23}=R_{12}/2=x_0/2$, respectively. 
In our calculation, we set the origin to be $({\bf r}_1+{\bf r}_2)/2$ and the
$x$-axis along the direction ${\bf r}_2-{\bf r}_1$, thus 
${\mathbf r}_{1,2}=(\mp x_0/2,0)$, ${\mathbf r}_3=(0,\sqrt{3}x_0/2)$ and $(0,0)$
for equilateral triangular and symmetric linear TQDs, respectively.

In the presence of external magnetic field 
$\mathbf{B}=B_{\perp}\hat{\mathbf{z}}+B_{\parallel}\hat{\mathbf{x}}$, 
the single-electron Hamiltonian reads\cite{lin2} 
\begin{eqnarray}
  H_{\rm e} = \frac{P_x^2+P_y^2}{2m_t}+V_{\rm c}({\mathbf r})+H_{\rm
    so}({\mathbf P})+H_{\rm Z}+H_{\rm E},
  \label{eq1}
\end{eqnarray}
where $\mathbf{P}=\mathbf{p}+(e/c)\mathbf{A}=-i\hbar
\mbox{\boldmath$\nabla$}+(e/c)\mathbf{A}$ 
with $\mathbf{A}=(-y,x)B_{\perp}/2$. 
It is noted that the orbital effect of the in-plane magnetic field is neglected 
due to a strong confinement along the $z$-axis.\cite{ka1}
$H_{\rm so}({\mathbf P})$ represents the SOC Hamiltonian including both 
the Rashba term\cite{Rashba1} due to the structure inversion asymmetry 
and the interface-inversion asymmetry (IIA) term.\cite{Ivchenko} 
Then it can be given by
\begin{eqnarray}
  H_{\rm so}({\mathbf P}) = a_0(P_x\sigma_y-P_y\sigma_x)+b_0(-P_x\sigma_x+P_y\sigma_y),\mbox{}
  \label{eq2}
\end{eqnarray}
with $\sigma_{x(y)}$, $a_0$ and $b_0$ representing the Pauli matrix, strengths
of the Rashba and IIA terms, respectively. 
$H_{\rm Z}=g\mu_{\rm B}(B_{\perp}\sigma_z+B_{\parallel}\sigma_x)/2$ is the Zeeman 
term where $g$, $\mu_{\rm B}$, and $\sigma_{z}$ stand for the Land{\'e}
factor,\cite{Graefi1} Bohr magneton,\cite{Gerlach} and Pauli matrix,
separately. 
$H_{\rm E}= eEx$ represents the electric field term with an electric field
applied along the $x$-direction.

To obtain a complete set of the single-electron orbital basis functions, 
which will be used to construct the three-electron ones, we define 
$H_{\rm 0}=(P_x^2+P_y^2)/(2m_t)+V_{\rm c}({\mathbf r})+H_{\rm E}$.
Due to the complex TQD confinement potential $V_{\rm c}({\mathbf r})$,
it is difficult to solve the Schr\"{o}dinger equation of $H_{\rm 0}$
analytically. 
Hence, we calculate the eigenvalues and eigenstates of $H_{\rm 0}$
numerically by employing the finite difference method according to the report by
Stano and Fabian.\cite{Fabian05}

For three-electron case, the total Hamiltonian can be expressed as 
\begin{eqnarray}
  H_{\rm tot} &=& H_{\rm e}^1+H_{\rm e}^2+H_{\rm e}^3+\mbox{}
  H_{\rm C}^{12}+H_{\rm C}^{23}+H_{\rm C}^{13}\mbox{}.
  \label{eq3}
\end{eqnarray}
Here, $H_{\rm e}^i$ ($i=1$-$3$) represents the single-electron Hamiltonian 
of the $i$th electron given by Eq.~(\ref{eq1}).
$H_{\rm C}^{ij}$ stands for the Coulomb interaction between the $i$th and 
$j$th electrons, which is given by
\begin{eqnarray}
  H_{\rm C}^{ij} = \frac{e^2}{4\pi\varepsilon_0\kappa|{\mathbf r}_i-{\mathbf r}_j|},
  \label{eq4}
\end{eqnarray}
with $\varepsilon_0$ and $\kappa$ being the vacuum dielectric constant
and relative static dielectric constant,\cite{Hada1} separately.

According to the approach widely used in the nuclear physics when 
building up the wavefunction of baryons composed of three quarks with
total spin being $1/2$ for each quark,\cite{Greiner,capstick} we
construct the three-electron basis functions as follows.
These three-electron basis functions are composed of both spin and orbital parts. 
We first construct the spin part in the form of either quartet or doublet\cite{DQ} 
by Clebsch-Gordan coefficient method.\cite{sakurai} 
It is noted that the quartet spin wavefunctions are exchange symmetric
whereas the doublet spin wavefunctions are of mixed symmetry.\cite{Richard} 
With these spin states, we then build up the corresponding orbital
wavefunctions to make the total three-electron basis functions exchange
antisymmetric for any two electrons.
The detailed expressions of three-electron total basis functions can be 
found in Ref.~\onlinecite{Zhe}.
These total basis functions are still denoted by either quartet
or doublet states according to their spin states. 

After obtaining these three-electron basis functions, one can calculate the
matrix elements of three-electron Hamiltonian $H_{\rm tot}$ [see Eq.~(3)]
including the orbital energy, Zeeman splitting, SOC and Coulomb
interaction.\cite{coulomb}
Since the matrix elements of the SOC is about three orders of magnitude smaller 
than others, we first calculate the three-electron eigenvalues and
eigenstates by exactly diagonalizing the three-electron Hamiltonian matrix in
the absence of SOC. Then based on the obtained three-electron eigenvalues and
eigenstates, we include the SOC perturbatively.

\section{NUMERICAL RESULTS}
In our calculation, the effective mass $m_t=0.19m_0$ with $m_0$ being the free
electron mass.\cite{Dexter} The strengths of the SOC are chosen 
as $a_0=-6.06\ $m/s and $b_0=-30.31\ $m/s corresponding to the electric field
along the growth direction being $30\ $kV/cm.\cite{Ivchenko}
The Land{\'e} factor $g=2$ (Ref.~\onlinecite{Graefi1}) and
the relative dielectric constant $\kappa=11.9$.\cite{Hada1} 
The grid points used to diagonalize the single-electron Hamiltonian are 
$95\times95$ with the accuracy of the energy being around $10^{-5}\ $meV. 
Based on these single-electron wavefunctions, the lowest 11194 doublet and 
10128 quartet basis functions are taken to obtain convergent three-electron 
energy spectra and eigenstates with the exact-diagonalization method. The
energy precision is also around $10^{-5}\ $meV. 

\subsection{Equilateral triangular TQDs}

In this part, we focus on the case of equilateral triangular TQDs. 
We first study both the perpendicular and parallel magnetic-field dependences of
three-electron energy spectra where detailed discussions about anticrossing
behavior induced by the equilateral triangular TQD confinement potential
and the SOC are presented. 
We then investigate the doublet-quartet transition of ground-state spin
configuration by tuning the interdot distance or the dot size in the
absence of external fields and the SOC. 
In addition, the electric-field dependence of energy levels and their charge
configurations are also shown.

\subsubsection{Anticrossings in the magnetic-field dependence of three-electron
  energy spectra}
\begin{figure}
  \begin{center}
    \includegraphics[width=8cm]{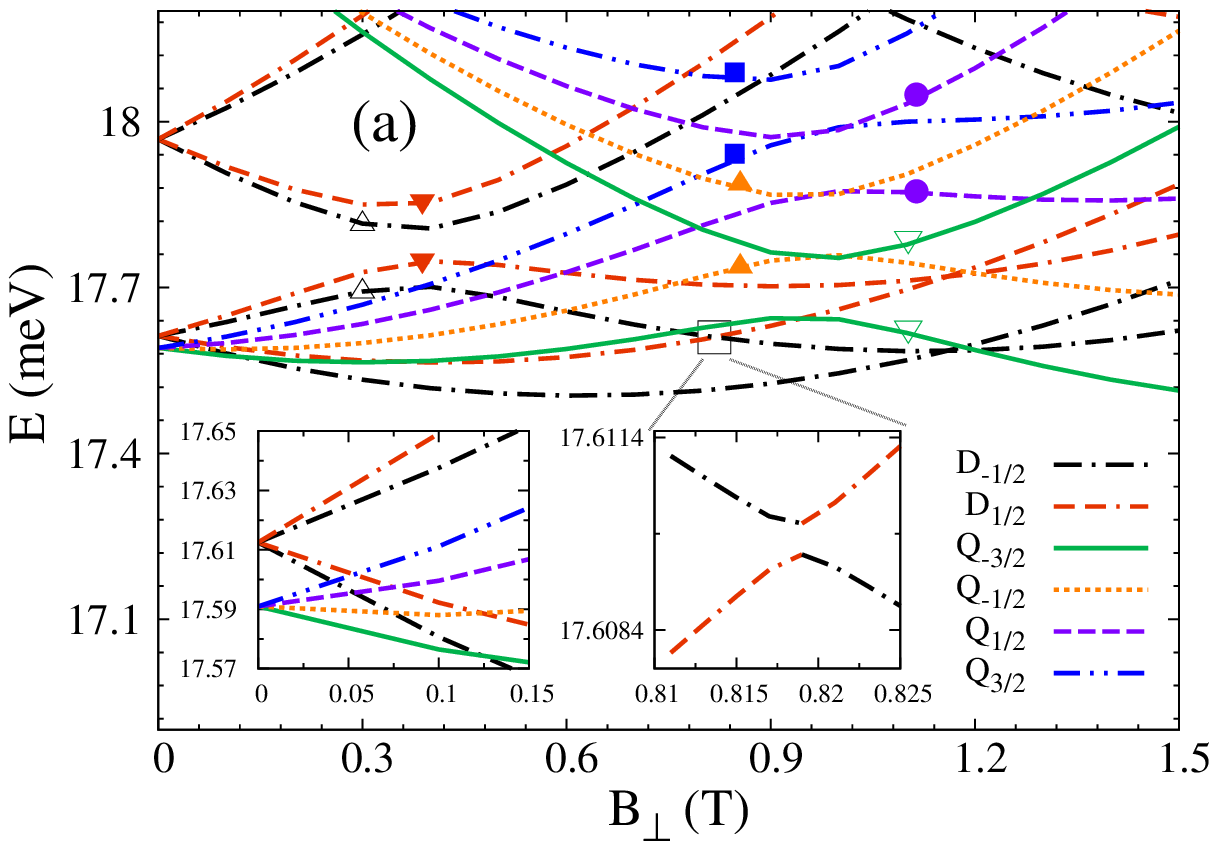} 
    \includegraphics[width=8.1cm]{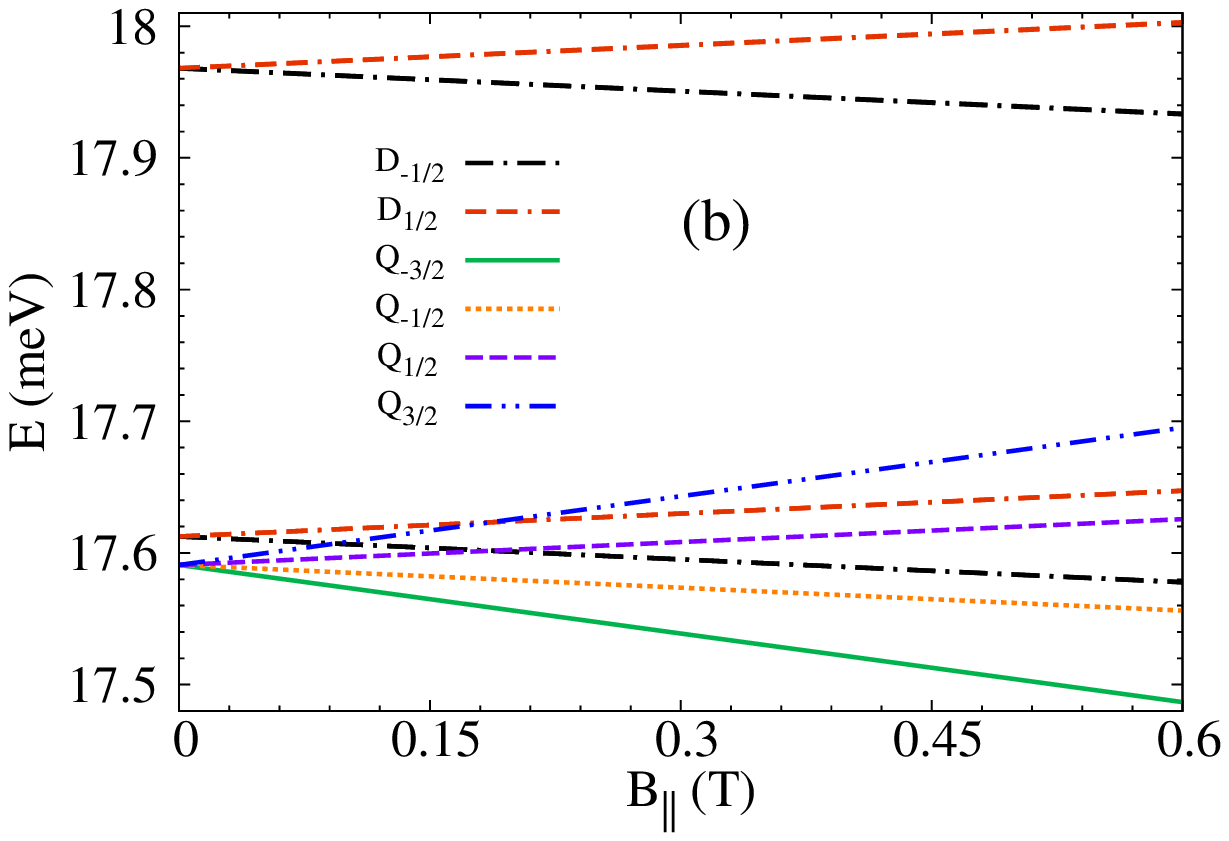} 
  \end{center}
  \caption{(Color online) (a) The lowest several energy levels {\it vs}. the perpendicular
    magnetic field $B_{\perp}$ in equilateral triangular TQDs. 
    The energy levels related to six anticrossings with large energy splittings
    are labelled by symbols $\blacktriangledown$, $\bigtriangleup$, $\bullet$,
    $\blacktriangle$, $\blacksquare$ and $\bigtriangledown$.
    The left inset enlarges the energy spectra in the vicinity of
    $B_{\perp}\sim 0\ $T whereas the right inset zooms one anticrossing induced
    by the SOC in the vicinity of $B_{\perp}\sim0.819\ $T.     
    (b) The lowest several energy levels {\it vs}. parallel magnetic field along the
    $x$-direction. Here, $d_0=29\ $nm and $x_0=11.6\ $nm.}
    \label{fig2}
\end{figure}

We first investigate the perpendicular magnetic-field dependence of energy
spectra in the absence of the SOC. 
In Fig.~\ref{fig2}(a), we plot the magnetic-field dependence of the lowest few
energy levels with the interdot distance $x_0$ being $11.6\ $nm and dot size
$d_0$ being $29\ $nm. These energy levels are denoted as either $D_{\pm1/2}$ or
$Q_{\pm1/2,\pm3/2}$ according to their spin states.
We find that, in the absence of the magnetic field, the ground state is 
four-fold degenerate quartet and the lowest doublet state is also four-fold
degenerate as shown in the left inset in Fig.~\ref{fig2}(a). 
With the magnetic field applied, the degeneracy of these states is lifted due to
the orbital effect of the magnetic field and the Zeeman splitting, leading to
many intersections.\cite{Zhe}
This is similar to the case of three electrons in single Si QDs.\cite{Zhe} 
In contrast to the single QD case, we find six
anticrossings with large energy splittings between the energy levels with the
same spin state in Fig.~\ref{fig2}(a). 
Specifically, at $B_{\perp}=0.4\ $T, we find two anticrossings with
energy splitting about $0.105\ $meV between two $D_{1/2}$ ($D_{-1/2}$)
  states labelled by $\blacktriangledown$ ($\bigtriangleup$). 
At $B_{\perp}=1\ $T, four anticrossings with energy splitting about
$0.102\ $meV are between the lowest two states of $Q_{S_z}$ with $S_z$ being
$\pm 1/2$ and $\pm3/2$ labelled by $\bullet$, $\blacktriangle$, $\blacksquare$
and $\bigtriangledown$, respectively. 
It is noted that this anticrossing behavior has not been reported in the
literature on TQDs.

The underlying physics of these anticrossings can be understood as follows. 
It is noted that, in the absence of the SOC, the total spin $S_{\rm tot}$ 
and its $z$-component $S_z$ are good quantum numbers. Therefore, the
Hilbert space can be divided into independent subspaces according to $S_{\rm tot}$ and $S_z$ 
and then we restrict to one subspace in the following.
For an infinitesimal interdot distance, i.e., $x_0 = 0\ $nm,
the equilateral triangular TQD confinement potential reduces to the limit of
single QD one. Due to the rotation symmetry of the single QD case, the total azimuthal
angular momentum $L$ is also a good quantum number. We plot the perpendicular
magnetic-field dependence of the lowest few energy levels with the
corresponding $L$ labelled in Fig.~\ref{fig6} in Appendix~\ref{appA}. We find
that there arise many intersections
between the energy levels with the same spin state but with different $L$,
which are just simple crossings. With the increase of interdot distance, the
single QD confinement potential  turns to the TQD one. The TQD confinement
potential deviates from the single 
QD one and breaks the rotation symmetry, which indicates that $L$ is no longer a
good quantum number in TQDs. Specifically, in our case with
equilateral triangular TQDs, the system has $C_3$ symmetry
instead.\cite{koster}
As a result, the subspace denoted by $S_{\rm tot}$ and $S_z$ can be
divided into three independent parts spanned by 
three-electron eigenstates in single QDs with
$\{L|L=3m+\chi\}$ ($m$ integral, $\chi=0,\pm1$), respectively.
It is noted that in each part, the eigenstates of three electrons in single QDs 
can be coupled with each other by the equilateral triangular
TQD confinement potential. 
This coupling can make the intersections between the states in the same part in
single QD case become anticrossing in TQD case as shown at $B_{\perp}=0.4\ $ and
$1\ $T in Fig.~\ref{fig2}(a). 

With the inclusion of the SOC, the three-electron energy spectra including the
large energy splittings of the above anticrossings are almost
unchanged. However, all the intersections mentioned 
previously become anticrossing with small energy splittings. 
One of these anticrossings (marked by open square at $B_{\perp}=0.819\ $T) 
is enlarged as shown in the right inset in Fig.~\ref{fig2}(a) where the energy 
splitting is about $0.49\ \mu$eV. 
This anticrossing behavior originates from the effect of the SOC with the
details shown in Appendix~\ref{appA}.

We then turn to study the effect of the parallel magnetic field applied along
the $x$-direction. In the absence of the SOC, the lowest several energy
levels as function of the magnetic field are shown in Fig.~\ref{fig2}(b) with
the same dot size and interdot distance as the case of perpendicular magnetic
field in Fig.~\ref{fig2}(a). 
These energy levels are denoted by either $D_{\pm 1/2}$ or $Q_{\pm 1/2,\pm 3/2}$
according to the spin quantum numbers $S_{\rm tot}$ and its $x$-component $S_x$. 
We find that in contrast to the complex perpendicular magnetic-field 
dependence of energy spectra shown in Fig.~\ref{fig2}(a), the energy levels show
a simple linear dependence on the parallel magnetic field here. 
This can be understood that the parallel magnetic field
affects the energy spectra only through the Zeeman 
splitting whereas the orbital effect is negligible owing to the
strong confinement 
along the growth direction. Specifically, for $D_{\pm 1/2}$ ($Q_{\pm 1/2,\pm 3/2}$), 
the $x$-components of the total spins are $\pm 1/2$ ($\pm 1/2,\pm 3/2$), which
indicate that these energy levels change linearly with the parallel magnetic field. 
This leads to the absence of the intersections between the energy levels
with the same spin state and therefore the anticrossings between these states
are absent. This is very different from the case of the perpendicular
magnetic field. In addition, similar to the perpendicular
magnetic field case,
we also find that all the intersections shown in Fig.~\ref{fig2}(b)
become anticrossing with the inclusion of the SOC.

\subsubsection{Doublet-quartet transition of ground-state spin configuration}

\begin{figure}
  \begin{center}
    \hspace{-0.4 cm}
    \includegraphics[width=8.cm]{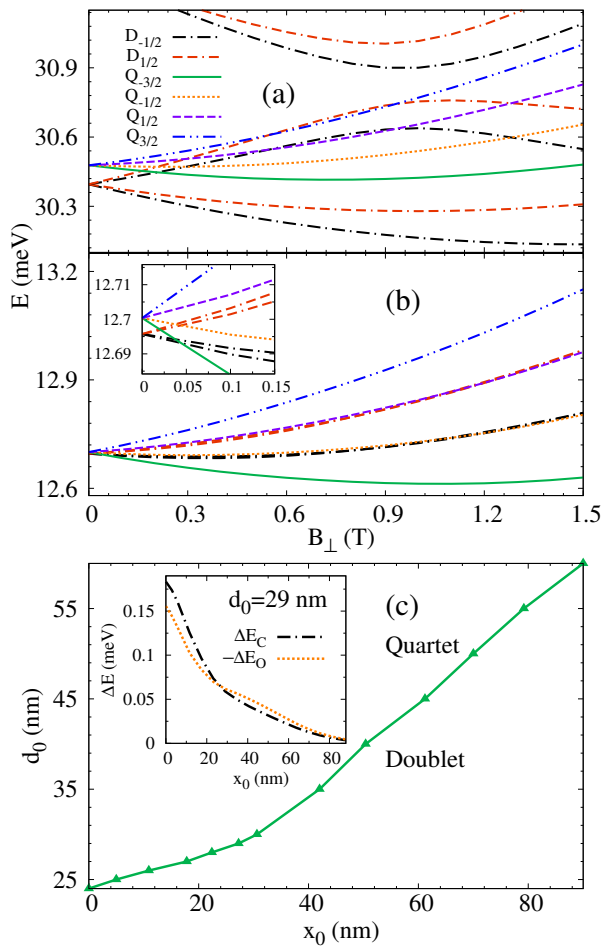} 
  \end{center}
  \caption{(Color online) The lowest few energy levels {\it vs}. perpendicular 
    magnetic field $B_{\perp}$ in equilateral triangular TQDs with (a) $d_0=20\
    $nm, $x_0=11.6\ $nm and (b) $d_0=29\ $nm, $x_0=58.0\ $nm. 
    The inset in (b) enlarges the energy spectra in the vicinity of
    $B_{\perp}\sim0$ T. 
    (c) Spin configuration of the ground state {\it vs}. dot size and interdot
    distance in the absence of external fields with
    the green solid curve with 
    symbol $\blacktriangle$ representing the crossover between different spin
    configurations ``Quartet'' and ``Doublet''. 
    In the inset, we also show the interdot-distance dependence of the Coulomb
    energy difference $\Delta E_{\rm C}$ (black chain) and orbital energy
    differences $-\Delta E_{\rm O}$ (yellow dotted) between the lowest doublet and
      quartet states. $d_0=29\ $nm.}
    \label{fig3}
\end{figure}

In addition to the anticrossing behavior, we also study the transition of
ground-state spin configuration in the absence of the external fields and the
SOC. 
It is noted that the ground state is quartet at zero magnetic field as shown in the 
left inset in Fig.~\ref{fig2} when $d_0=29\ $nm and $x_0=11.6\ $nm. This spin
configuration of the ground state in the absence of external fields 
has not been reported in the literature on TQDs. 
By varying either the dot size ($d_0=20\ $nm, $x_0=11.6\ $nm) or interdot distance 
($d_0=29\ $nm, $x_0=58.0\ $nm), we find that the ground-state spin
configurations become doublet instead at $B_{\perp}=0\ $T as shown in
Figs.~\ref{fig3}(a) and (b), respectively. Therefore, the
doublet-quartet transition of ground-state spin configuration can be realized by
tuning either the dot size or the interdot distance. 
Furthermore, to show the spin configuration of ground state as function of the dot
size and the interdot distance, we plot a phase-diagram-like picture in 
Fig.~\ref{fig3}(c) by calculating the energy difference $\Delta E_{\rm DQ}$
between the lowest doublet ($E_{\rm D}$) and quartet ($E_{\rm Q}$) states (i.e.,
$\Delta E_{\rm DQ}=E_{\rm D}-E_{\rm Q}$). 
In this figure, the doublet and quartet ground-state spin configurations are 
separated by a solid curve with $\blacktriangle$ corresponding to 
$\Delta E_{\rm DQ}=0$. Specifically, as the dot size decreases, the ground-state
spin configuration changes from quartet to doublet, which is similar to the case
in single QDs reported by Liu {\em et al.}.\cite{Zhe}

As for the effect of interdot distance $x_0$, we also find a transition of the
spin configuration of ground state from quartet to doublet as $x_0$ increases. 
This can be understood as follows. 
It is noted that the energy difference between the lowest doublet and quartet
states $\Delta E_{\rm DQ}$ is contributed by the orbital energy difference
$\Delta E_{\rm O}$ and Coulomb energy difference $\Delta E_{\rm C}$, i.e.,
$\Delta E_{\rm DQ}=\Delta E_{\rm C}+\Delta E_{\rm O}$.  
With a specific dot size $d_0$ being $29\ $nm, we calculate the
interdot-distance dependence of $\Delta E_{\rm C}$ and $-\Delta E_{\rm O}$
as shown in Fig.~\ref{fig3}(c). 
We find that both $\Delta E_{\rm C}$ and $-\Delta E_{\rm O}$ decrease with the
increase of interdot distance. However, $\Delta E_{\rm C}$ decreases faster than
$-\Delta E_{\rm O}$ and an intersection (i.e., $\Delta E_{\rm DQ}=0$) occurs at
$x_0 \sim 26\ $nm, indicating a transition of ground-state spin configuration
from quartet to doublet. 
This is similar to the case of double QDs where the transition of ground-state 
spin configuration can also be realized by tuning the interdot
distance.\cite{lin2}

\subsubsection{Electric-field dependence of energy spectra and charge
  configurations}
\begin{figure}
  \begin{center}
    \includegraphics[width=8cm]{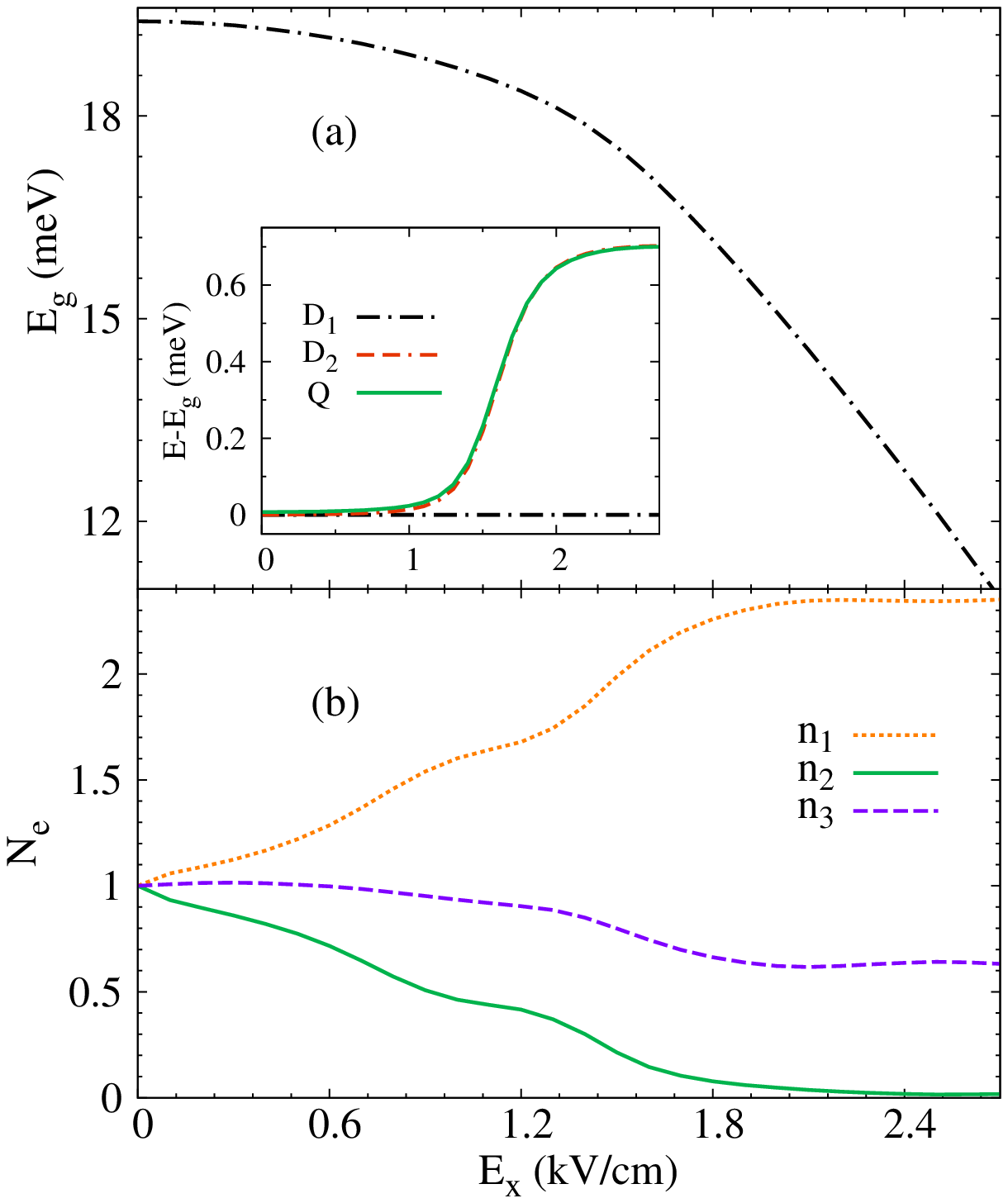} 
    \includegraphics[width=7cm]{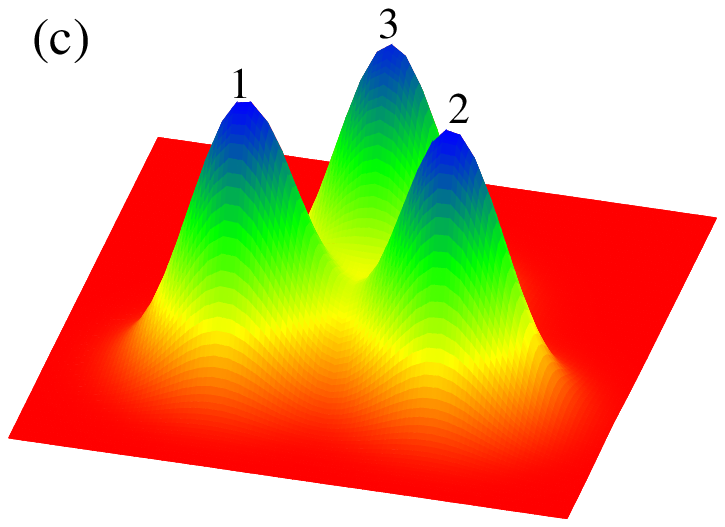} 
    \includegraphics[width=7cm]{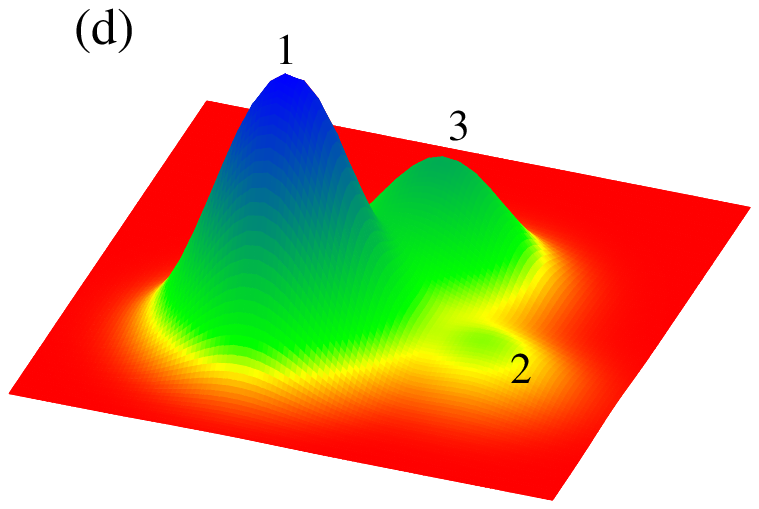} 
  \end{center}
  \caption{(Color online) (a) Energy of the ground state $E_{\rm g}$ {\it vs}. 
    electric field applied along the $x$-direction $E_x$. 
    The lowest several energy levels with the energy of the ground state
    subtracted are shown in the inset.
    $D_1$ and $D_2$ are two-fold degenerate doublet states whereas $Q$
    represents the lowest four-fold degenerate quartet states. 
    (b) Charge configuration ($n_1,n_2,n_3$) of ground state {\it vs}. $E_x$.
    (c) and (d) show the charge density distributions of
    three electrons in the ground
    states with $(1,1,1)$ and $(2,0,1)$ configurations, respectively.
    Here, $d_0=20\ $nm and $x_0=58\ $nm.}
    \label{fig4}
\end{figure}
We also investigate the dependence of the lowest several energy
levels on the electric field along $x$-direction in the absence of the SOC and
magnetic field. 
We first focus on the ground state as shown in Fig.~\ref{fig4}(a)
which is two-fold degenerate doublet.
We find that the ground-state energy changes slowly in small electric 
field regime whereas it decreases rapidly when the electric field becomes
strong. 
This can be understood from the variation of charge configuration of the ground 
state. 
As shown in Fig.~\ref{fig4}(b), the charge configuration is denoted by
$(n_1,n_2,n_3)$ with $n_i$ ($i=1$-$3$) representing the electron occupation
number of the $i$th QD.\cite{sp-conf}
It is seen that, in small electric field regime, the charge configuration is 
close to $(1,1,1)$, whereas it changes to the configuration close to $(2,0,1)$
when the electric field is strong. 
The charge density distribution of three 
electrons in $(1,1,1)$ and $(2,0,1)$ 
configurations are plotted in Figs.~\ref{fig4}(c) and (d), respectively. 
It is noted that the electric field term is given by $H_{\rm E}= eEx$ according
to Eq.~(\ref{eq1}). 
As a result, the variation of the single-electron energy in the middle QD (i.e.,
dot 3) is negligible since dot 3 is located at $x=0$. 
In contrast, the single-electron energy in the right QD (i.e., dot 2) is raised
whereas that in the left QD (i.e., dot 1) is suppressed due to the applied
electric field.  
Therefore, close to $(1,1,1)$ configuration, i.e., the small electric field regime,
the net contribution of the electric field is relatively small and thus the energy 
changes slowly with the increase of the electric field. 
When the electric field becomes strong, i.e., close to $(2,0,1)$ configuration, the
energy due to the electric field decreases with increasing electric field
whereas the contribution of the Coulomb energy presents an opposite trend. It is
noted that the variation of the energy induced by the electric field is much
larger than that of the Coulomb energy, which leads to a rapid decrease of the
ground-state energy with the increase of the electric field.

Then we turn to the lowest several excited states. 
The dependence of these energy levels on the electric field is plotted in the
inset in Fig.~\ref{fig4}(a). 
In this figure, the energy levels are shown with the energy of the ground state
subtracted to make them distinguishable.
It is noted that the excited doublet state is two-fold degenerate whereas
the excited quartet state is four-fold.
Moreover, the behavior of the charge configurations of these states as function
of the electric field is similar to that of the ground state.

\subsection{Symmetric linear TQDs}
\begin{figure}
  \begin{center}
    \includegraphics[width=8cm]{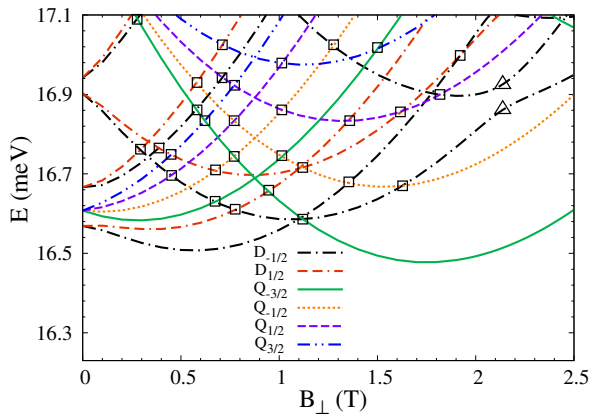} 
  \end{center}
  \caption{(Color online) 
    The lowest several energy levels {\it vs}. the perpendicular magnetic field
    $B_{\perp}$ in symmetric linear TQDs. The energy levels related to the
    anticrossing with large energy splitting are labelled by $\bigtriangleup$.
    Open squares denote the intersections which become anticrossing due to the
    SOC. $d_0=29\ $nm and $x_0/2=11.6\ $nm.}
    \label{fig5}
\end{figure}

In this part, we investigate the case of symmetric linear TQDs where the perpendicular
magnetic-field dependence of the lowest several energy levels with $x_0/2=11.6\
$nm and $d_0=29\ $nm in the absence of the SOC are shown in Fig.~\ref{fig5}. 
These energy levels are still denoted as either $D_{\pm1/2}$ or $Q_{\pm1/2,\pm3/2}$
according to their spin states. It is seen that, as the magnetic field
increases, there arise many intersections due to the Zeeman splitting and the
orbital effect of the magnetic field, which is similar to the equilateral
triangular TQD case. Additionally, at $B_{\perp}=2.2\ $T, one also observes an
anticrossing between the lowest two states of $D_{-1/2}$ labelled by
$\bigtriangleup$. The underlying physics can be understood similar to the case
of the equilateral triangular TQDs.
We also focus on a specific subspace denoted by $S_{\rm tot}$ and $S_z$. 
This subspace can be further divided into two independent parts since 
the system with the symmetric linear TQD confinement potential has $C_2$
symmetry.\cite{koster} These two parts are spanned by the three-electron eigenstates
in single QDs with $\{L|L=2m+\chi\}$ ($m$ integral, $\chi=0,1$), respectively.
The intersection between the energy levels in the same part in single QDs 
is lifted to show anticrossing
behavior in symmetric linear TQD case as shown at $B_{\perp}=2.2\ $T. Moreover,
we also find that only part of intersections show anticrossing behavior with the
inclusion of the SOC, which are labelled by open squares in Fig.~\ref{fig5}. The
details are shown in Appendix~\ref{appA}. This is different from the equilateral
triangular TQD case where all the intersections become anticrossing. 

In addition, compared with the equilateral triangular TQD
case, similar behaviors such as the doublet-quartet transition of ground-state
spin configuration by varying the dot size or interdot distance in the absence
of the magnetic field and the SOC, variation of three-electron charge configuration by the
electric field, and the linear dependence of energy spectra on parallel magnetic
field can also be observed in symmetric linear TQDs. 

\section{SUMMARY}
In summary, we have investigated the three-electron energy spectra in laterally 
coupled SiGe/Si/SiGe TQDs with single valley approximation by utilizing the
real-space configuration interaction method. 
The electron-electron Coulomb interaction, which is crucial to the energy
spectra, is explicitly included whereas the relatively small SOC is treated
perturbatively. 
The dependences of the energy spectra on the dot size, 
interdot distance, (either perpendicular or parallel) magnetic field, and
electric field are studied in both the equilateral triangular and symmetric
linear TQD cases.
In both cases, we find doublet-quartet transitions of ground-state spin
configurations by varying either the dot size or interdot distance in the
absence of
external fields, which has not been reported in the literature on
TQDs. Interestingly, we also observe anticrossings with large energy splittings
between energy levels with the same spin state in the perpendicular 
magnetic-field dependence of the energy spectra in the absence of the SOC. 
These anticrossings, which have not been reported in the literature,
originate from the equilateral triangular and symmetric linear TQD confinement
potential, respectively. 
In contrast to the complex dependence on the perpendicular magnetic field, 
the energy spectra vary linearly with the parallel magnetic field due to
the negligible orbital effect of parallel magnetic field. Additionally, in
perpendicular/parallel magnetic-field dependence of energy levels, we find that
all the intersections in equilateral triangular TQD cases become anticrossing
due to the SOC whereas only part of them show anticrossing behaviors in symmetric
linear TQD case. All these anticrossing behaviors are analyzed from the symmetry
consideration. 
Moreover, we also find that energy levels and their charge
configurations can be strongly affected by the in-plane electric field. 

\begin{acknowledgments}
This work was supported by the National Basic Research 
Program of China under Grant No.\ 2012CB922002 
and the Strategic Priority Research Program of the Chinese 
Academy of Sciences under Grant No. XDB01000000. 
One of the authors (YFR) would like to thank Y. Yin and M. Q. Weng
for valuable discussions in numerical calculation. 

\end{acknowledgments}

\begin{appendix}
\section{ANTICROSSINGS DUE TO THE SOC IN TQDS}\label{appA}
\begin{figure}
  \begin{center}
    \includegraphics[width=8cm]{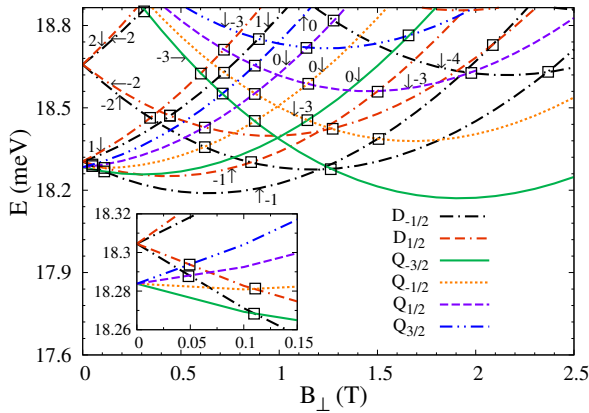} 
  \end{center}
  \caption{(Color online) The lowest few energy levels {\it vs}. perpendicular magnetic
    field in single QDs. 
    The total azimuthal angular momentum $L$ of each energy
    level is labelled in the figure. Open squares denote the intersections
    which can become anticrossing due to the SOC. The inset enlarges the energy
    spectra in the vicinity of $B_{\perp}\sim 0\ $T. $d_0=29\ $nm.}
    \label{fig6}
\end{figure}
In this part, we show the detailed analysis about the anticrossing behavior due
to the SOC in the magnetic field dependence of the energy spectra 
in both equilateral triangular and symmetric linear TQD cases. We 
focus on the perpendicular magnetic field case.
To facilitate the understanding, we first review the anticrossing behavior in
single QD limit.\cite{Zhe} 
In the absence of the SOC, the perpendicular magnetic-field dependence of the
lowest few energy levels are plotted in Fig.~\ref{fig6}. 
These energy levels are denoted by the good quantum numbers $S_{\rm tot}$, $S_z$
and $L$, and the intersections between them are simple crossings as mentioned
previously.\cite{Zhe} 
With the inclusion of the SOC, part of these intersections (labelled by
open squares in Fig.~\ref{fig6}) are lifted to show anticrossing behavior. 
To understand this behavior, we span the Hilbert space by the three-electron
eigenstates labelled by ($L,S_{\rm tot},S_z$). It is noted that the Hilbert space
is also spanned by these eigenstates in the following. 
As the eigenstate with ($L,S_z$) can be coupled with the one with
($L\pm1,S_z\mp1$) or ($L\pm1,S_z\pm1$) by the SOC [see Eq.~(\ref{eq2})]
according to the previous works,\cite{lin1,Zhe} the Hilbert space can be divided
into two independent parts spanned by the eigenstates with $L+S_z+3/2$ being
even or odd. The eigenstates in the same part can be coupled with each other by
the SOC, which makes the intersections between the energy levels in the same
part become anticrossing.

With the increase of the interdot distance, we turn to study the equilateral
triangular TQD case.
As pointed out previously, the equilateral triangular TQD confinement potential 
can couple the eigenstates in single QDs with the same spin state and the
difference of $L$ being $3m$ ($m$ integer). This indicates that two independent
parts (i.e., $L+S_z+3/2$ being even or odd) can be coupled by the TQD
confinement potential. 
As a result, all the intersections in equilateral triangular TQD case can be
lifted by the SOC.

As for the symmetric linear TQD case, the confinement potential can couple 
the eigenstates in single QDs with the same spin state and the difference of $L$
being even. Thus, two independent parts (i.e., $L+S_z+3/2$ being even or odd)
are still independent, which is different from the equilateral triangular TQD
case. As a result, only the intersections between the energy levels in the same
part can show anticrossing behavior due to the SOC. 

We also study the case of parallel magnetic field applied along $x$-axis. 
Similar to the case of perpendicular magnetic field, in single QD limit, the
three-electron eigenstates can be divided into two independent parts according
to $L+S_x+3/2$ being either even or odd with the inclusion of the
SOC.\cite{Zhe,lin1} As for the TQD cases, we also find that all the intersections in 
equilateral triangular TQDs whereas only those between the energy levels in the
same part (i.e., $L+S_x+3/2$ being either even or odd) in symmetric linear TQDs
can become anticrossing due to the SOC.
\end{appendix}


\begin{thebibliography}{0}
\bibitem{DQ} For quartet states, the total spin $S_{\rm tot}=3/2$ and its
  component parallel to the external magnetic field along the $z$($x$)-direction is
  $S_{z(x)}=\pm 1/2,~ \pm3/2$. As for the doublet states, $S_{\rm tot}=1/2$ and
  $S_{z(x)}=\pm 1/2$.
\bibitem{sp-conf} $n_i$ can be calculated by 
$n_i=3 \int_{\Omega_i} {\mathrm d} {\mathbf r}_1\iint {\mathrm d} {\mathbf r}_2 
{\mathrm d} {\mathbf r}_3 |\Psi({\mathbf r}_1,{\mathbf r}_2,{\mathbf r}_3)|^2$,
where $\Psi$ is the wavefunction of three-electron eigenstate. 
The region of integration for the first integral is $\Omega_i$ in which the
position is nearest to the center of the $i$th QD, whereas the region is the
whole two dimensional real space for the second and third ones.
\end{thebibliography}
\end{document}